\documentclass[graybox]{svmult}

\usepackage{mathptmx}       % selects Times Roman as basic font
\usepackage{helvet}         % selects Helvetica as sans-serif font
\usepackage{courier}        % selects Courier as typewriter font
\usepackage{type1cm}        % activate if the above 3 fonts are
\usepackage{makeidx}         % allows index generation
\usepackage{graphicx}        % standard LaTeX graphics tool
\usepackage{multicol}        % used for the two-column index
\usepackage[bottom]{footmisc}% places footnotes at page bottom

\makeindex             % used for the subject index
\usepackage{amsmath}
\usepackage{amssymb}
\usepackage{amsfonts}

\begin{document}

\title*{Efficient hedging in Bates model using high-order compact
  finite differences}
%\titlerunning{Efficient hedging in Bates model}
\author{Bertram D\"uring and Alexander Pitkin}
\institute{Bertram D\"uring \at Department of Mathematics, University
  of Sussex, Pevensey II, Brighton, BN1 9QH, UK, \email{bd80@sussex.ac.uk}
\and Alexander Pitkin \at Department of Mathematics, University
  of Sussex, Pevensey II, Brighton, BN1 9QH, UK, \email{a.h.pitkin@sussex.ac.uk}}
\maketitle

\abstract{We evaluate the hedging performance of a high-order compact
  finite difference scheme from \cite{DuPi17} for option pricing in
  Bates model.
We compare the scheme's hedging performance to standard finite
difference methods in different examples. We observe that the new scheme
outperforms a standard, second-order central
finite difference approximation in all our experiments. 
}

\section{Introduction}
\label{sec:1}

The Bates model \cite{Bates} can be considered as the market standard in
financial option pricing applications. It combines the positive
features of stochastic volatility and jump-diffusion models.
In this model the option price
is given as the solution of a partial integro-differential equation (PIDE),
see e.g.\ \cite{Cont}. 

In \cite{DuPi17} we have presented a new high-order compact finite
difference scheme for option pricing in Bates model. The
implicit-explicit scheme is based on the approaches in D\"uring and
Fourni\'e \cite{DuFo12} and Salmi {\em et al.\/}
\cite{Salmi14}. The scheme is fourth order accurate in space and second order accurate
in time. It requires only one initial $LU$-factorisation of a
sparse matrix to perform the option price valuation. Due to its
structural similarities with standard second-order finite difference
schemes it can be employed to to upgrade existing implementations in a
straightforward manner to obtain a highly efficient option pricing
code.

In the present work we evaluate the hedging performance of the scheme
derived in \cite{DuPi17}.
We compare the scheme's hedging performance to standard finite difference methods where the new scheme
outperforms a standard discretisation, based on a second-order central
finite difference approximation, in all our experiments. 

This article is organised as follows. In the next section we recall Bates
model for option pricing and the related
partial integro-differential equation. We refer here to the \cite{DuPi17} paper for the derivation of the {\em implicit-explicit high-order compact finite difference scheme\/} which we adapt and implement to conduct the numerical experiments. Section~\ref{sec:3} 
is devoted to the computation of the so-called {\em Greeks\/} and
the evaluation of the scheme's hedging performance in two examples of hedged portfolios.

\section{The Bates Model}
\label{sec:2}
% Always give a unique label
% and use \ref{<label>} for cross-references
% and \cite{<label>} for bibliographic references
% use \sectionmark{}
% to alter or adjust the section heading in the running head
%Instead of simply listing headings of different levels we recommend to
%let every heading be followed by at least a short passage of text.
%Further on please use the \LaTeX\ automatism for all your
%cross-references and citations.
%
%Please note that the first line of text that follows a heading is not indented, whereas the first lines of all subsequent paragraphs are.
%
%
%Use the standard \verb|equation| environment to typeset your equations, e.g.
%%
%\begin{equation}
%a \times b = c\;,
%\end{equation}
%%
%however, for multiline equations we recommend to use the \verb|eqnarray| environment\footnote{In physics texts please activate the class option \texttt{vecphys} to depict your vectors in \textbf{\itshape boldface-italic} type - as is customary for a wide range of physical subjects}.
%\begin{eqnarray}
%a \times b = c \nonumber\\
%\vec{a} \cdot \vec{b}=\vec{c}
%\label{eq:01}
%\end{eqnarray}

The Bates model \cite{Bates} is a stochastic volatility model which allows for jumps in returns. Within this model the behaviour of the asset value, \textit{S}, and its variance, \( \sigma \), is described by the coupled stochastic differential equations,
\begin{align*}
dS(t) &= \mu_B S(t) dt + \sqrt{\sigma (t)} S(t) dW_1(t) + S(t) dJ,\\
d\sigma(t) &= \kappa(\theta - \sigma(t)) + v\sqrt{\sigma (t)} dW_2(t),
\end{align*}
for $ 0\leqslant t \leqslant T $ and with $ S(0), \sigma(0) > 0  $. Here, $ \mu_B = r - \lambda\xi_B $ is the drift rate, where $r\geqslant0$ is the risk-free interest rate. The jump process $ J $ is a compound Poisson process with intensity $ \lambda\geqslant0$ and $J+1$ has a log-normal distribution $p(\tilde{y})$ with the mean in $\log (\tilde{y})$ being $\gamma$ and the variance in $\log (\tilde{y})$ being $v^2$, i.e.\ the probability density function is given by
 \[p(\tilde{y})=\frac{1}{\sqrt{2\pi}\tilde{y}v}e^{-\frac{(\log\tilde{y}-\gamma)^2}{2v^2}}.\]
The parameter $\xi_B$ is defined by $\xi_B=e^{\gamma+\frac{v^2}{2}}-1$. The variance has mean level $\theta$, $\kappa$ is the rate of reversion back to mean level of $\sigma$ and $v$ is the volatility of the variance $\sigma$. The two Wiener processes $W_1$ and $W_2$ have correlation $\rho$.

\subsection{Partial Integro-Differential Equation}
By standard derivative pricing arguments for the Bates model, we
obtain the partial integro-differential equation
\begin{eqnarray*}
\frac{\partial V}{\partial t} +
  \frac{1}{2}S^2\sigma\frac{\partial^2 V}{\partial S^2}+\rho v\sigma
  S\frac{\partial^2 V}{\partial S \partial \sigma}
  +\frac{1}{2}v^2\sigma \frac{\partial^2 V}{\partial \sigma^2} +
  (r-\lambda\xi_B)S\frac{\partial V}{\partial S} + \kappa(\theta -
  \sigma) \frac{\partial V}{\partial \sigma}\\ - (r+\lambda)V +
  \lambda \int_0^{+\infty} \! V(S\tilde{y},v,t)p(\tilde{y}) \,
  \mathrm{d}\tilde{y} = L_D V+L_I V,
\end{eqnarray*}
which has to be solved for $S,\sigma > 0$, $0 \leq t < T $ and subject
to a suitable final condition, e.g.\ $V(S,\sigma,T) = \max(K-S,0), $
in the case of a European put option, with $K$ denoting the strike price.  
For clarity the operators $L_D V$ and $L_I V$ are defined as the differential part (including the term $-(r+\lambda)V$) and the integral part, respectively.

Through the following transformation of variables
\begin{equation*} x=\log S , \quad \tau = T-t , \quad   y=\frac{\sigma}{v} \quad \text{and} \quad u= \exp (r+\lambda)V  \end{equation*}
we obtain
% \begin{multline*} 
% \label{eq:Ptransf}
% u_\tau = \frac{1}{2}vy\left(\frac{\partial^2 u
%     }{\partial x^2}+\frac{\partial^2 u}{\partial y^2}\right)+\rho
%   vy\frac{\partial^2 u}{\partial x \partial y}
%   -\left(\frac{1}{2}vy-r+\lambda\xi_B\right)\frac{\partial u}{\partial
%     x} \\+ \kappa \frac{(\theta - vy) }{v}\frac{\partial u}{\partial y}
%  +\exp (r+\lambda)L_I V,
% \end{multline*}
\begin{multline*} 
\label{eq:Ptransf}
u_\tau = \frac{1}{2}vy\left(\frac{\partial^2 u
    }{\partial x^2}+\frac{\partial^2 u}{\partial y^2}\right)+\rho
  vy\frac{\partial^2 u}{\partial x \partial y}
  -\left(\frac{1}{2}vy-r+\lambda\xi_B\right)\frac{\partial u}{\partial
    x} \\+ \kappa \frac{(\theta - vy) }{v}\frac{\partial u}{\partial y}
 \lambda \int^{+\infty}_{-\infty} \tilde u(x + z , y , \tau ) \tilde p(z) \, \mathrm{d}z,
\end{multline*}
which is now posed on $ \mathbb{R} \times \mathbb{R} ^+ \times (0,T), $ with
 % \begin{equation*} L_I V = \lambda \int_0^\infty \! V(S\tilde{y},v,t)p(\tilde{y}) \, \mathrm{d}\tilde{y}. \end{equation*}
% Applying the same transformation to the intergral term, $L_I$,
%  \begin{equation*}\exp (r+\lambda)L_I V = \lambda \int_0^{+\infty}  u(x \tilde y , y , \tau ) p(\tilde y) \, \mathrm{d}\tilde y .\end{equation*}
% Now by setting $ z=\log \tilde y,$ $\tilde u (z,y,\tau) = u(e^z,y,\tau) $ and $ \tilde  p (z) = e^z p(e^z) $, we have
%  \begin{equation*} \exp (r+\lambda)L_I V = \lambda \int_0^{+\infty}
%  u(x \tilde y , y , \tau ) p(\tilde y) \, \mathrm{d}\tilde y  =
%  \lambda \int^{+\infty}_{-\infty} \tilde u(x + z , y , \tau ) \tilde
%  p(z) \, \mathrm{d}z. \end{equation*}
$\tilde u (z,y,\tau) = u(e^z,y,\tau) $ and $ \tilde  p (z) = e^z p(e^z) $.
The problem is completed by suitable initial and boundary conditions, which for a European put option are: 
\begin{align*}
 u(x,y ,0)&= \max (1- \exp(x),0), \quad x \in \mathbb{R}, \; y > 0, \\
 u(x,y,t) &\rightarrow 1, \quad x \rightarrow - \infty , \; y  > 0, \;
            t > 0, \\
  u(x,y,t) &\rightarrow 0, \quad x \rightarrow + \infty , \; y > 0, \; t > 0,  \\
u_y (x,y,t) &\rightarrow 0, \quad x \in \mathbb{R} , \; y \rightarrow \infty , \; t > 0, \\
 u_y(x,y,t) &\rightarrow 0, \quad x \in \mathbb{R} , \;
  y  \rightarrow 0, \; t > 0. 
\end{align*}

\subsection{Implicit-explicit high-order compact scheme}
\label{subsec:2}
%Instead of simply listing headings of different levels we recommend to
%let every heading be followed by at least a short passage of text.
%Further on please use the \LaTeX\ automatism for all your
%cross-references\index{cross-references} and citations\index{citations}
%as has already been described in Sect.~\ref{sec:2}.
%
%\begin{quotation}
%Please do not use quotation marks when quoting texts! Simply use the \verb|quotation| environment -- it will automatically render Springer's preferred layout.
%\end{quotation}

For the discretisation, we replace
$\mathbb{R}$ by $[-R_1,R_1]$ and $\mathbb{R}^+$ by $[L_2,R_2]$ with
$R_1,R_2 >L_2>0$. We consider a uniform grid $ Z = \{ x_i \in
[-R_1,R_1] : x_i = i h_1 ,\; i =-N, ...  ,N\} \times \{\sigma_j \in
[L_2,R_2] : \sigma_j = L_2 + j h_2 ,\; j=0, ... , M\} $ consisting of
$(2N+1) \times (M+1) $ grid points with $R_1 = N h_1$ , $R_2 = L_2 + M
h_2 $ and with space step $h:=h_1=h_2$ and time step $ k $. Let $
u_{i,j}^n $ denote the approximate solution of (2) in $(x_i, \sigma_j)
$ at the time $t_n = n k$ and let $u^n = (u_{i,j}^n) $.  

For the numerical solution of the partial integro-differential
equation we use the implicit-explicit high-order compact scheme
presented in \cite{DuPi17}. The implicit-explicit discretisation in time
is accomplished through an adaptation of the Crank-Nicholson method
which includes an explicit treatment for the integral operator. The
scheme is fourth-order accurate in space and second-order accurate in
time. We refer to \cite{DuPi17} for
the details of the derivation of the scheme and the implementation of
the initial and boundary conditions.  

If not mentioned otherwise, we use the following default parameters in
our numerical experiments: $\kappa=2$, $\theta=0.01$, $\rho=-0.5$,
$\nu=0.1$, 
$r=0.05$,
$\lambda=0.2$,
$\gamma=-0.5$.

\section{The Greeks}
\label{sec:3}

The so-called {\em Greeks\/} are the partial derivatives of the option price with respect to
independent variables or parameters. These quantities represent 
the market sensitivities of options.
Practitioners use these quantities to gain an insight into the effects of different market conditions on an options price and furthermore to develop hedging strategies against unfavourable changes in a portfolio of assets. 

\subsection{Vega}

Vega measures the sensitivity of
the option price with respect to changes in the volatility of the underlying asset, i.e.\ 
\[ \text{Vega} = \frac{\partial{V}}{\partial{\sigma}}. \]

Vega represents the amount that an options price changes in reaction to a $1\%$ change in the implied volatility of the underlying asset. We examine whether the higher-order convergence achieved in the option price will also be represented in the vega of the option. 

We calculate vega directly from the option price $u(x,y,t)$. The order of the scheme is maintained by using following fourth-order approximation formula, while the boundaries are trimmed to remove the need for extrapolation,
\[ \text{Vega}_{i,j}^n = \frac{1}{\sigma_j} \frac{ u_{i-2,j}^n -8u_{i-1,j}^n +8u_{i+1,j}^n -u_{i+2,j}^n
      }{12h}. \]
\begin{figure}[b]
\sidecaption
% For example, with the graphicx style use
\includegraphics[scale=.45]{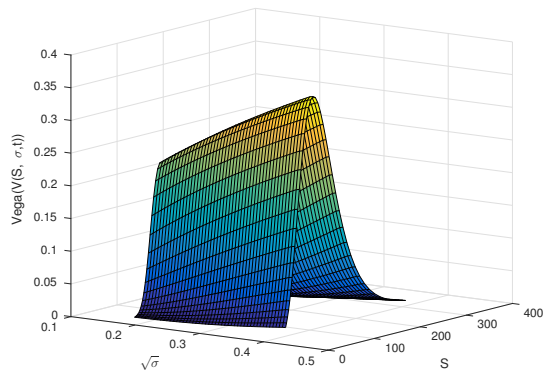}
%
% If no graphics program available, insert a blank space i.e. use
%\picplace{5cm}{2cm} % Give the correct figure height and width in cm
%
%\caption{If the width of the figure is less than 7.8 cm use the \texttt{sidecapion} command to flush the caption on the left side of the page. If the figure is positioned at the top of the page, align the sidecaption with the top of the figure -- to achieve this you simply need to use the optional argument \texttt{[t]} with the \texttt{sidecaption} command}
\caption{Vega of European put option priced under the Bates model with parameters: Strike $K=100$, time to expiry $T=0.5$. }
\label{fig:1}       % Give a unique label
\end{figure}
We conduct a numerical study to evaluate the rate of convergence of vega. We refer to both the $ l_2 $-norm error $\epsilon_2$ and the $ l_{\infty} $-norm error $\epsilon_\infty $ with respect to a numerical reference solution on a fine grid with $h_{\text{ref}} =0.025. $ By fixing the parabolic mesh ratio $k/h^2$ we expect these errors to converge as $ \epsilon= Ch^m $ for some constants $m$ and $C$. We generate a double-logarithmic plot of $\epsilon$ against $h$ which should be asymptotic to a straight line with slope $m$, with $m$ being the experimentally determined order of the scheme. 

\begin{figure}[b]
\sidecaption
\includegraphics[scale=.47]{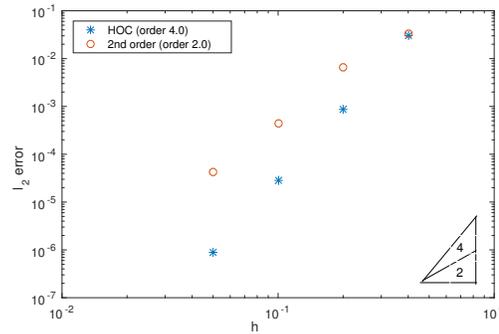}
\caption{Convergence of $l_2$-error of the vega of a European put option priced under the Bates model with parameters: Strike $K=100$, time to expiry $T=0.5$. }
\label{fig:2}       % Give a unique label
\end{figure}

\begin{figure}[b]
\sidecaption
\includegraphics[scale=.47]{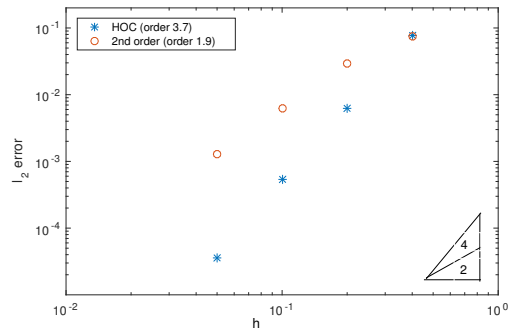}
\caption{Convergence of $l_{\infty}$-error of the vega of a European put option priced under the Bates model with parameters: Strike $K=100$, time to expiry $T=0.5$. }
\label{fig:3}       % Give a unique label
\end{figure}

As a tool for comparison we perform the same numerical study using a
standard second-order central difference scheme. The results of these experiments are seen in  Fig.~\ref{fig:2} and   Fig.~\ref{fig:3}.
We observe here that the experimentally determined convergence rates
match well the theoretical order of each scheme. The errors at coarse
grid, $h=0.4$, are comparable, while on finer grids the high-order
compact scheme gives orders of magnitude better accuracy on the same
grids, achieving convergence rates of about fourth order.

\subsection{Hedging Vega}

As with all financial trading, options are subject to risk and managing this risk is key to success. One method of managing risk is to establish a hedge against the implied volatility of the underlying asset. This is achieved by creating a vega neutral option position, which will be not be sensitive to fluctuations in volatility.

% Hedges can be made against the risks of price sensitivity, second-order time price sensitivity and time sensitivity.

\subsubsection{Hedging Example 1}

\begin{figure}[b]
\sidecaption
% For example, with the graphicx style use
\includegraphics[scale=.41]{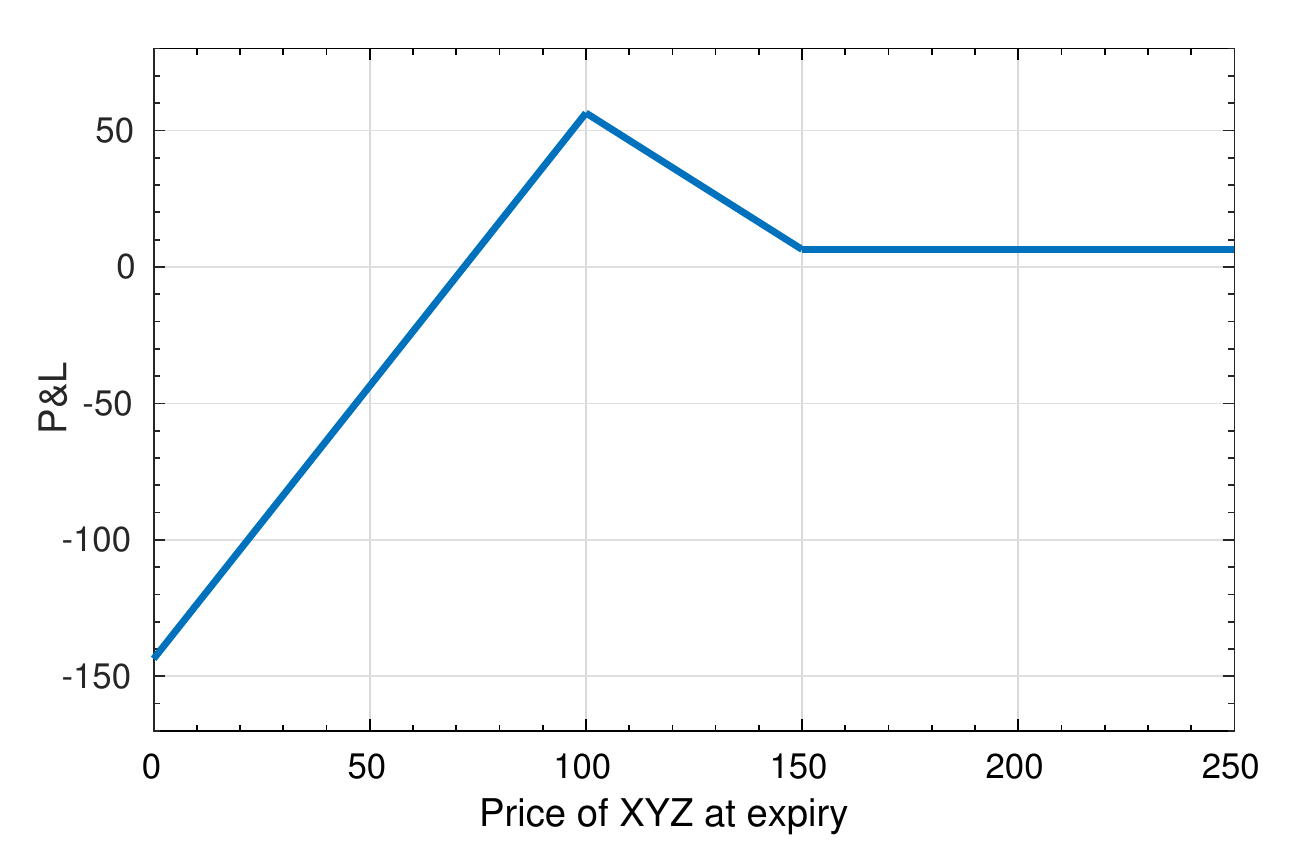}
%
% If no graphics program available, insert a blank space i.e. use
%\picplace{5cm}{2cm} % Give the correct figure height and width in cm
%
%\caption{If the width of the figure is less than 7.8 cm use the \texttt{sidecapion} command to flush the caption on the left side of the page. If the figure is positioned at the top of the page, align the sidecaption with the top of the figure -- to achieve this you simply need to use the optional argument \texttt{[t]} with the \texttt{sidecaption} command}
\caption{Payoff for ratio vertical put spread, examples include a 1:2 spread, where the trader writes two put options then goes long one put option with a higher strike price.}
\label{fig:4}       % Give a unique label
\end{figure}

An investment fund holds a long position in a non dividend paying stock, XYZ, which is currently trading at \$135. The investment fund wishes to secure an income from the position and writes some put options for XYZ with strike \$100. The investment fund now has a position with negative vega. To hedge this vega risk the investment fund creates a ratio vertical put spread by buying put options with strike \$150, creating a payoff diagram as shown in  Fig.~\ref{fig:4}.

We propose that using the high-order compact (HOC) scheme the investment fund can utilise the high-order convergence in vega to achieve a more accurate vega hedge when constructing the ratio spread. To measure this we compare the ratio used for each mesh size, $h$, with the fine reference grid and examine the resulting percentage error.

The results for the high-order scheme and those for a comparative second-order scheme are shown in Table.~\ref{tab:1}. The high-order scheme significantly outperforms the second-order scheme at all mesh-sizes, suggesting that when entering a large position the HOC scheme will lead to a significant improvement in the vega hedge.

\begin{table}
\caption{Percentage error in vega hedge ratio}
\label{tab:1}       % Give a unique label
%
% Follow this input for your own table layout
%
\begin{tabular}{p{1.5cm}p{1.7cm}p{2.2cm}p{2cm}p{1.7cm}p{2cm}}
\hline\noalign{\smallskip}
Scheme & Mesh-size & Percentage error & Scheme & Mesh-size & Percentage error  \\
\noalign{\smallskip}\svhline\noalign{\smallskip}
HOC & 0.4  & 33.3138 & Second-order & 0.4 & 62.0312 \\
HOC & 0.2 & 6.7519 & Second-order & 0.2 & 33.0638 \\
HOC & 0.1 & 0.6251  & Second-order & 0.1 & 7.4073 \\
HOC & 0.05 & 0.0400  & Second-order & 0.05 & 1.5364 \\
\noalign{\smallskip}\hline\noalign{\smallskip}
\end{tabular}
%$^a$ Table foot note (with superscript)
\end{table}
%

\subsection{Gamma}

Gamma is the second derivative of the option price with respect to the underlying asset. Gamma measures the rate of change in an option's delta, providing information on the convexity of the option's value in relation to the price of the underlying asset,
\[ \Gamma = \frac{\partial^2{V}}{\partial{S}^2}. \]

\begin{figure}[b]
\sidecaption
% For example, with the graphicx style use
\includegraphics[scale=.30]{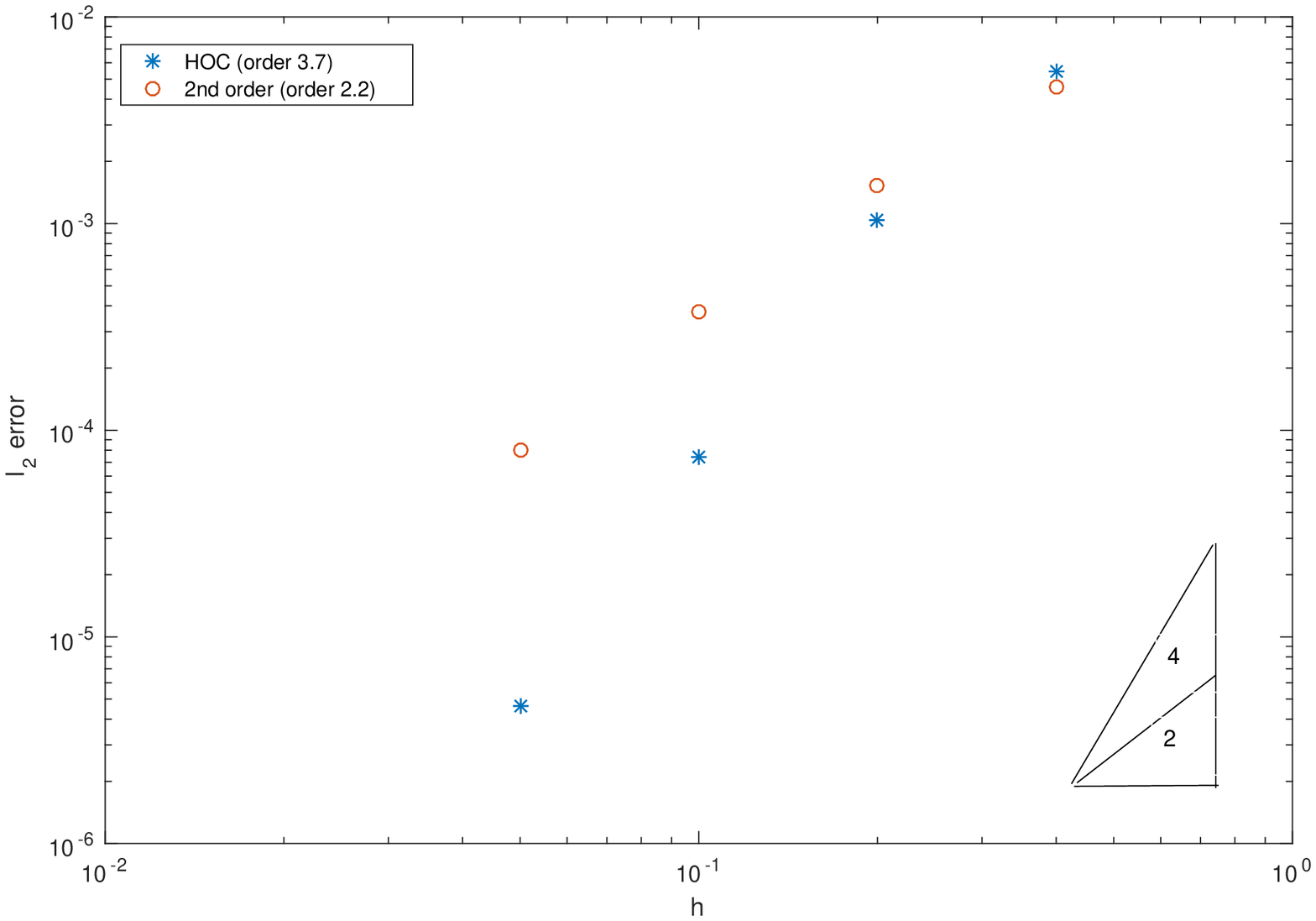}
%
% If no graphics program available, insert a blank space i.e. use
%\picplace{5cm}{2cm} % Give the correct figure height and width in cm
%
%\caption{If the width of the figure is less than 7.8 cm use the \texttt{sidecapion} command to flush the caption on the left side of the page. If the figure is positioned at the top of the page, align the sidecaption with the top of the figure -- to achieve this you simply need to use the optional argument \texttt{[t]} with the \texttt{sidecaption} command}
\caption{Convergence of $l_2$-error of gamma of a European put option priced under the Bates model with parameters: Strike $K=100$, time to expiry $T=0.5$.}
\label{fig:6}       % Give a unique label
\end{figure}

\begin{figure}[b]
\sidecaption
% For example, with the graphicx style use
\includegraphics[scale=.29]{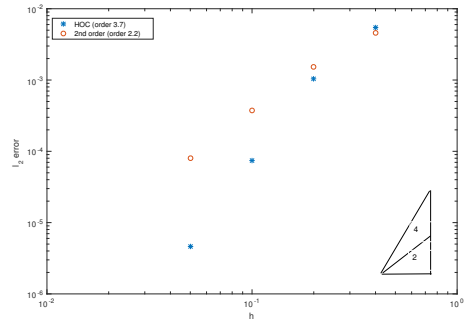}
%
% If no graphics program available, insert a blank space i.e. use
%\picplace{5cm}{2cm} % Give the correct figure height and width in cm
%
%\caption{If the width of the figure is less than 7.8 cm use the \texttt{sidecapion} command to flush the caption on the left side of the page. If the figure is positioned at the top of the page, align the sidecaption with the top of the figure -- to achieve this you simply need to use the optional argument \texttt{[t]} with the \texttt{sidecaption} command}
\caption{Convergence of $l_{\infty}$-error of gamma of a European put option priced under the Bates model with parameters: Strike $K=100$, time to expiry $T=0.5$.}
\label{fig:7}       % Give a unique label
\end{figure}

We calculate gamma directly from the option price $u(x,y,t)$. To maintain the order of the scheme we use the following fourth-order approximation formula, the boundaries are trimmed to remove the need for extrapolation,
\[ \Gamma_{i,j}^n = \frac{1}{S_i^2} \frac{ u_{i-2,j}^n -16u_{i-1,j}^n +30u_{i,j}^n -16u_{i+1,j}^n +u_{i+2,j}^n}{12h^2}. \]
\begin{figure}[t]
\sidecaption[t]
\includegraphics[scale=.30]{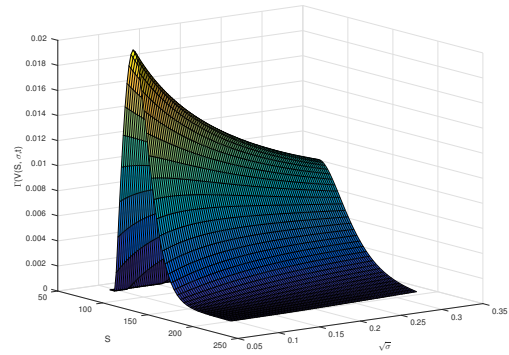}
%
% If no graphics program available, insert a blank space i.e. use
%\picplace{5cm}{2cm} % Give the correct figure height and width in cm
\caption{Gamma of European put option priced under the Bates model with parameters: Strike $K=100$, time to expiry $T=0.5$. }
\label{fig:5}       % Give a unique label
\end{figure}
We conduct a numerical study to evaluate the rate of convergence of gamma. We refer to both the $ l_2 $-error $\epsilon_2$ and the $ l_{\infty} $-error $\epsilon_\infty $ with respect to a numerical reference solution on a fine grid with $h_{\text{ref}} =0.025. $ For comparison we perform the same numerical study using a standard second-order central difference scheme. The results of these experiments are seen in  Fig.~\ref{fig:6} and   Fig.~\ref{fig:7}.

The high-order compact scheme achieves convergence rates between three and four for the $l_2$- and $l_{\infty}$-errors, respectively. This is an improvement on the second-order scheme and suggests that the high-order scheme is beneficial when developing trading strategies which involve a gamma hedge.

\subsection{Hedging Gamma}

Hedges of gamma risk are often accompanied by a delta hedge, with delta being the first derivative of the option price with respect to the underlying asset. A delta hedged portfolio is not subject to risk owing to a change in the price of the underlying asset, the gamma hedge is a re-adjustment of this delta hedge.

Delta-gamma hedging strategies often require frequent adjustments and hence are subject to high trading costs. However, if executed correctly they can enable the holder to exploit positions with positive theta, meaning the position is profitable over short time durations.

\subsubsection{Hedging Example 2}

An analyst at an investment fund looks to create a strategy with positive theta against the funds currently held assets. They choose a ratio write spread, which involves writing options at a higher strike price than they are purchased. The analyst is wary of the positions risk related to move in the underlying asset and hence adjusts the ratio of short to long options to eliminate the net gamma.

The resulting position will have a delta value which must be hedged before the analyst can assess any profitability from the positive theta of the spread. The delta of the two option positions long and short is totalled and if positive or negative underlying assets are sold or bought, respectively.

The resulting theta is calculated and if positive the analyst can recommend the strategy as a short term trade for the investment fund. 

We propose that using the high-order compact (HOC) scheme the investment fund can utilise the high-order convergence in gamma to achieve a more accurate gamma hedge ratio. To measure this we compare the ratio used for each mesh size, $h$, with the fine reference grid and examine the resulting percentage error.

The results for the high-order scheme and those for a comparative second-order scheme are shown in Table.~\ref{tab:2}. The high-order scheme offers better results at all mesh-sizes, this improvement is particularly important in hedged positions which require repeat computation and regular adjustments.

\begin{table}
\caption{Percentage error in gamma hedge ratio}
\label{tab:2}       % Give a unique label
%
% Follow this input for your own table layout
%
\begin{tabular}{p{1.5cm}p{1.7cm}p{2.2cm}p{2cm}p{1.7cm}p{2cm}}
\hline\noalign{\smallskip}
Scheme & Mesh-size & Percentage error & Scheme & Mesh-size & Percentage error  \\
\noalign{\smallskip}\svhline\noalign{\smallskip}
HOC & 0.4  & 14.8885 & Second-order & 0.4 & 25.1112 \\
HOC & 0.2 & 2.3323 & Second-order & 0.2 & 6.3482 \\
HOC & 0.1 & 0.1281  & Second-order & 0.1 & 1.3304 \\
HOC & 0.05 & 0.0081  & Second-order & 0.05 & 0.2674 \\
\noalign{\smallskip}\hline\noalign{\smallskip}
\end{tabular}
%$^a$ Table foot note (with superscript)
\end{table}
%

% \section{Conclusions}
% \label{sec:conc}

% Using the new high-order compact finite difference method for
% option pricing in stochastic volatility jump models developed by [CITEDURINGPITKIN]. We
% have conducted numerical experiments to confirm high-order convergence in both the vega
% and gamma of the option. These experiments are followed by examples of trading strategies, where 
% gains in convergence and computation time may be utilised by market participants.

\begin{acknowledgement}
BD acknowledges partial support by the Leverhulme Trust research project grant `Novel discretisations for higher-order nonlinear PDE' (RPG-2015-69). 
AP has been supported by a studentship under the EPSRC Doctoral
Training Partnership (DTP) scheme (grant number EP/M506667/1).
\end{acknowledgement}

\end{document}